\documentstyle[aps,multicol]{revtex}

\def\gs{\gtrsim}
\def\ls{\lesssim}

\begin{document}
\draft
\bibliographystyle{prsty}
\title{Kinetic Heterogeneities in a Highly Supercooled Liquid}
\author{Ryoichi Yamamoto and Akira Onuki}
\address{Department of Physics, Kyoto University, Kyoto 606-01}
\date{Published in {\it J. Phys. Soc. Jpn.} {\bf 66}, 2545 (1997)}
\maketitle

\begin{abstract}
We study  a highly supercooled two-dimensional
 fluid mixture via molecular dynamics  simulation. 
 We follow bond breakage events among particle pairs, 
which   occur on the  scale of 
 the $\alpha$ relaxation time $\tau_{\alpha}$. 
 Large scale heterogeneities analogous to the critical 
 fluctuations in Ising systems are found 
 in the spatial distribution of bonds which are broken in 
 a time interval with a width of order $0.05\tau_{\alpha}$.  
The structure factor of the broken bond 
 density is well approximated by the Ornstein-Zernike form.
The correlation length is  of order  
 $100 \sigma_1$ at  the lowest temperature studied, 
$\sigma_1$ being the particle size.
The weakly bonded regions thus identified 
 evolve  in time with  strong spatial correlations.  
\end{abstract}
\pacs{Keywords: glass transition, supercooled liquid, 
 heterogeneities, bond breakage}

\begin{multicols}{2}

As fluids are supercooled below their  melting temperature,
 their structural relaxations  become very slow,
 resulting in a glass transition at $T=T_g$ characterized by 
 a dramatic increase of  the viscosity.  \cite{Jackel,Ediger} 
Recently  much  attention has been paid to 
the mode coupling theory, \cite{mode1,mode2} 
 which  is a self-consistent 
 decoupling  scheme for the density time correlation 
 function and describes  onset of 
 glassy slowing down or slow structural relaxations 
 considerably above $T_g$.  
 The dominant fluctuation contribution arises from   the 
 first peak wave number of the structure factor.  
For a long time, however, 
 it has been intuitively expected  
 \cite{Adam,Cohen,Stillinger,Perera} that 
  dynamics in glasses  should be cooperative, 
 involving many molecules,  owing to configuration 
 restrictions. 
  Adam and Gibbs \cite{Adam} speculated  that particle 
 motions over the interparticle distance or the potential barrier 
 can take place only collectively 
 in {\it cooperatively rearranging regions} (CRR)
 and  such regions have  a minimum size 
 which  grows  as the temperature is lowered.
On the other hand, 
the mode coupling theory does not predict long range 
correlations.

In  his recent   experiments 
 Fischer \cite{Fischer} has reported   large excess light scattering 
 with a correlation length $\xi$ ($20-200$ nm) 
 which increases on  approaching 
 the glass transition from a liquid state. 
This indicates the presence of large scale heterogeneities in 
 supercooled  liquids.  
To examine this effect, some authors used 
  Monte Carlo simulations 
 on a dense polymer melt. \cite{Ray,Weber}   
 There,  significant dependence of the diffusion 
constant on the system size was found, 
indicating correlations 
up to the system size. \cite{Ray}  
Very recently,  Weber {\it et al}\cite{Weber}  have found 
  that short range nematic 
orientational order  produces  enhancement of 
long range density correlations.  
 The work of Weber {\it et al}  suggests that  large scale 
density heterogeneities will  not be  produced in simple molecular 
systems.

In their  MD simulation on a simple two-dimensional fluid 
 in a deeply supercooled state, Muranaka and Hiwatari 
 \cite{Muranaka} visualized  significant 
large scale heterogeneities   in particle displacements in a
 relatively short time interval which corresponds to the
 so-called $\beta$ relaxation.
In liquid states with higher temperatures, 
 Hurley and Harrowell \cite{Hurley} 
 observed   similar kinetic heterogeneities   
 but the correlation length was still 
 on the order of a few particle diameters.
 The characterization of  
 these patterns has not been made in these papers.

It is a natural expectation that structural changes of 
the  configurations of particles  occur collectively \cite{Adam}, 
but it is  rather surprising that the visualization of 
such long range correlations has been nonexistent in 
the structural or $\alpha$ relaxation. 
The aims  here  are to visualize them unambiguously 
and  analyze the patterns qualitatively.

Our system is composed of two different atomic species,  
 $1$ and $2$, with the  numbers $N_{1}=N_{2}=5000$, 
 interacting via the soft-core potential 
 $v_{\alpha\beta}(r)=
 \epsilon [(\sigma_{\alpha}+\sigma_{\beta})/2r]^{12}$, where 
 $r$ is the distance between two particles and $\alpha, \beta= 1,2$.
We take  the size and mass ratios as 
  $\sigma_{2}/\sigma_{1}=1.4$ and  
 $m_{2}/m_{1}=2$. 
The  difference 
between  the particle sizes prevents crystallization. 
 There is no enhancement of 
the composition fluctuations at any wave number  and the system is 
effectively one-component.
The interaction is truncated at $r =4.5\sigma_{1}$.
The leapfrog algorithm\cite{Allen} is adopted to solve the differential
 equations with a time step of $0.005\tau$ under the periodic 
 boundary condition.
The space and time are measured in units of $\sigma_1$ and  
 $\tau_0=({m_{1}\sigma_{1}^{2}/\epsilon})^{1/2}$.
The temperature is kept at a constant value using the Gaussian 
 constraint thermostat,\cite{Allen}  and the density is 
 fixed at $n=0.8/\sigma_{1}^{2}$.
We specify the thermodynamic states 
 using the  coupling parameter 
 $\Gamma_{eff}= n(\epsilon/k_{\rm B}T)^{1/6}\sum_{\alpha,\beta} 
x_{\alpha}x_{\beta}(\sigma_{\alpha} + \sigma_{\beta})^2/4$, 
\cite{Muranaka,Bernu}  where
 $x_1$ and $x_2$ are the compositions and are equal to 
$1/2$ in our case, so it follows that $\sigma_{eff} \cong 1.21\sigma_1$
Data are taken  at 
 $\Gamma_{eff}=1.0,\ 1.1,\ 1.2,\ 1.3$ and $1.4$.
The corresponding 
temperature is $2.54,\ 1.43,\ 0.850,\ 0.526$ and $0.337$, 
 respectively, in units of $\epsilon/k_{\rm B}$.
The corresponding 
pressure is $32.8,\ 27.0,\ 23.6,\ 21.63$  and $20.5$,
 respectively, in units of $\epsilon/\sigma_1^2$. 
We  realized these states by cooling 
 the system very slowly in a stepwise manner
 from a liquid state with $\Gamma_{eff}=0.8$.
The equilibration time is $5000$ after  $\Gamma_{eff}$ is changed 
 from $1.3$ to $1.4$. There was no appreciable change in  
 the pressure 
over  computation times of order $10000$.
The  zero-shear viscosity $\eta$ is of order
 $6$ at $\Gamma_{eff}=1$  and  $10^4$ 
 at  $\Gamma_{eff}=1.4$ in units of 
$m_1/\tau_0$ \cite{rheo}.  It was claimed  that glassy behavior becomes 
apparent for $\Gamma_{eff} \gs 1.3$. \cite{Muranaka_private}

In our analysis we follow bond breakage processes. 
For each atomic configuration given at time $t_{0}$, 
 a pair of atoms $i$ and $j$ is considered to be bonded if 
 \begin{equation}
 r_{ij}(t_{0})= |{\bf r}_{i}(t_{0})-{\bf r}_{j}(t_{0})|\leq 1.1 
\sigma_{\alpha\beta},
 \label{1}
 \end{equation}
 where $i$ and $j$ belong to the species $\alpha$ and $\beta$,
 respectively, and $\sigma_{\alpha\beta}= 
{\frac{1}{2}}(\sigma_{\alpha}+\sigma_{\beta})$.
 In our case the particle density is so high 
 that  the pair distribution 
functions $g_{\alpha\beta}(r)$ have  a sharp peak at 
$r \cong \sigma_{\alpha\beta}$. 
Thus the  definition of bonds is very insensitive 
 to the factor in front of 
$\sigma_{\alpha\beta}$ as long as it is 
 slightly  larger than 1. 
 The  polygons, which are 
  composed of the bonds and cover the space,  are mostly 
triangles as shown in Fig. 1.    After a lapse of time  
$t$ we  count the  number $N_{suv}(t)$ of 
 surviving bonds which continue to satisfy 
 \begin{equation} 
 r_{ij}(t_{0}+t)\leq 1.6\sigma_{\alpha\beta}.
 \label{2}
 \end{equation}
If the initially bonded pair  does not satisfy (2), we regard it 
 to have been broken. This definition of bond breakage is also 
insensitive to the factor  
in front of $\sigma_{\alpha\beta}$ as long as it is 
 considerably   larger than 1 and smaller than 2. 
We have confirmed that $N_{suv}(t)$ decays exponentially as  
 \begin{equation}
 N_{suv}(t)=N_{suv}(0)\exp\left(-{t}/{\tau_{b}}\right).
 \label{3}
 \end{equation}
This holds excellently in the whole time region
 for $\Gamma_{eff} \leq 1.2$, 
 but has been obtained from the fit in the short time region 
 $t \ls \tau_b$ for  $\Gamma_{eff} \geq 1.3$. 
We have not yet checked  whether or not    
$N_{suv}(t)$ relaxes  exponentially for $t \gs \tau_{b}$. 
Note that there are three kinds of bonds in our two-component fluid, but 
 no significant differences can be found in their breakage processes.
In Fig. 2 we show that the bond breakage time 
 $\tau_{b}$ grows strongly with increasing $\Gamma_{eff}$. 
 Our data  may  be well   fitted to 
$\tau_b  \propto T^{-4}$  as well as to  
$\log \tau_b \propto 1/T$.  
Because  $\tau_b$  is the relaxation 
time of the glassy structures, 
  $\tau_b$ should be of 
 the same order as the $\alpha$ relaxation time $\tau_{\alpha}$. 
In fact  
 for the same model Muranaka and Hiwatari \cite{Muranaka_private} 
 obtained  $\tau_{\alpha}$  on the same order as our 
$\tau_b$  from the self part of 
the time correlation function $F_s(q,t)$ at $q=2\pi/\sigma_1$. 
This coincidence  is natural because 
 $F_s(q,t)$ at $q=2\pi/\sigma_1$
 decays when a tagged particle moves 
  over the interparticle distance.

Fig. 3  displays the  bonds broken within the time 
 interval $[t_0,t_0+ 0.05\tau_b]$ at $\Gamma_{eff}=1.0$ and $1.3$.
 The center positions  
 ${\bf R}_{ij}= {\frac{1}{2}}({\bf r}_{i}(t_{0})+{\bf r}_{j}(t_{0}))$
 at the initial time $t_0$ of the pairs  broken within the time 
 interval are depicted here.  
For the glassy case $\Gamma_{eff}=1.3$,  
 the broken bond distribution is markedly  heterogeneous, 
  being composed of 
 clusters with various sizes, whereas  for the liquid case
 $\Gamma_{eff}=1$,  the inhomogeneity is much weaker. 
More precisely, in the  bond breakage  
we  observe strings of broken bonds involving   several particles 
  even in liquid states,   because of the high density of our system,    
  and  their large-scale  aggregation  in glassy states. 
The origin of the heterogeneities is  ascribed 
 to the fact that  
 particle motions over the interparticle distance 
 can occur only collectively in glassy states. \cite{Adam}  
Furthermore, in Fig. 4 we show the broken bonds  
 in  two consecutive time intervals,  $[t_0,t_0+ 0.05\tau_b]$ and 
 $[t_0+0.05\tau_b,t_0+ 0.1\tau_b]$,  at $\Gamma_{eff}=1.4$.
The initial times at which the bonds are defined are $t_0$ and 
 $t_0+ 0.05\tau_b$ for the two cases.
We can see that the clusters of  broken bonds in the two time 
 intervals  mostly overlap or are adjacent to each other.
This means that {\it weakly bonded regions} or {\it relatively 
 active regions} 
migrate  in space  on the time scale of $\tau_{\alpha}$.
The $\alpha$ relaxation ends when most of the bonds are broken.

We next calculated the structure factor $S_b(q)$ 
 of the broken bond density
 $\rho_{b}({\bf r})=\sum \delta({\bf r}-{\bf R}_{ij})$  defined by
 \begin{equation}
 S_b(q)=N_b^{-1} {\bigg \langle} {\bigg |} \sum 
 \exp(i{\bf q}\cdot{\bf R}_{ij})    
 {\bigg |}^{2} {\bigg \rangle}_{a}.
 \label{4}
\end{equation}
The summation is over the broken pairs, $N_b$ is the total number 
 of the broken bonds,  and ${\langle \cdots \rangle}_{a}$ 
 is the angular  average over the direction of the wave vector 
 ${\bf q} = ({2\pi}/{L})(n_x,n_y)$ where
 $n_x,n_y=\pm1,\pm2,\pm3.\cdots$.
The system length $L$ is $111.8$ here. 
Furthermore, since $S_b(q)$ from one configuration  fluctuates  
at small $q$, we have taken the averages of $S_b(q)$ from 10 
sequential  configurations  for $\Gamma_{eff}=1.0-1.3$ 
and  4 sequential configurations for $\Gamma_{eff}=1.4$.  Then, 
as shown in Fig.5, $S_b(q)$ can  be well   approximated by 
 the simple Ornstein-Zernike form, 
 \begin{equation}
 S_b(q)={S_b(0)}/[1+\xi^{2}q^{2}] ,
 \label{5}
 \end{equation}
 from which  we can determine $\xi$.
In addition, the raw $S_b(q)$ from (4) tends to become 
 1 at large $q$, so it  has been subtracted 
 in  $S_b(q)$  in Fig. 5.
(This  subtracted value is negligibly small for 
 $\Gamma_{eff}>1.1$ at long wavelengths $q\xi <1$, however.) 
Interestingly, 
\begin{equation}
S_b(0) \propto \xi^2
\label{6}
\end{equation}
 and the large $q$ behavior of 
 $S_b(q)$ is insensitive to $\Gamma_{eff}$ as in Ising spin systems 
 near the critical point. 
In Fig.6 we show $\xi$ vs $1/T$  for the 
 two  widths of the time intervals, 
 $0.05\tau_b$ ($\Diamond$) and $0.1\tau_b$ ($\bullet$).
We can see that $\xi$ is largely  independent of the width except 
 for the lowest temperature case, where $\Gamma_{eff}=1.4$.
In this case, however, $\xi$ approaches  the system length 
 $L=111.8$ and a finite size effect should be present.
Also $\tau_b$ at the lowest $T$ in Fig. 2 seems to be affected 
 by the same effect.

Figures 3-5 suggest that the cluster structure in the highly supercooled 
 states comprises clusters with various  length scales,
 the minimum being $\sigma_1$ and the maximum being $\xi$. 
It appears to be  self-similar if we consider  clusters  with 
 size $\ell$ in the region 
 $\sigma_1 \ll \ell \ll  \xi$  
 in the limit  $\xi \gg \sigma_1$,
 which is also supported by the power law behavior $S_b(q) \sim 1/q^2$ 
 shown in Fig. 5. 
This apparently contradicts Adam-Gibbs' phenomenological 
 speculation  in which  the {\it minimum}  size of  CRR grows upon cooling 
 \cite{Adam}. 
We stress that the heterogeneities are of purely kinetic origin and 
 are dynamical objects.
Indeed, the small clusters with $\ell \ll \xi$ in Figs. 3 and 4 
 evolve into larger ones or dissolve  in the subsequent 
 time intervals.

We have also calculated the structure factors 
 of the density, composition and stress tensor, but  no 
 indication of  large scale heterogeneities has been detected in such 
 static quantities in accordance  with previous papers.\cite{Dasgupta,Ernst}
 The kinetic heterogeneities found
 in this paper play a crucial   role in response to externally applied 
  perturbations such as sound waves or shear flow. \cite{rheo}

A frequently disputed  issue in the literature is whether or not 
 there is an underlying thermodynamic phase transition at a nonzero 
 temperature $T_0$ below $T_g$ in highly supercooled liquids.
In our case,  Fig. 6 does not  indicate any  divergence of $\xi$ at a nonzero 
 temperature, although this is not conclusive due to the finite size 
 effect arising from $\xi \sim L$.
We  mention a recent simulation on a three-dimensional Lennard-Jones 
 mixture by  Ghosh and Dasgupta, \cite{Ghosh} 
 who  used scaling relations of finite systems 
 to conclude  $\xi \propto T^{-1.5}$ as $T \rightarrow 0$ 
 without examining cluster structures. 
The limited data in Fig. 6 also yield 
 $|\partial \log \xi/\partial \log T| \sim 1$.

In summary, by performing long time and large scale 
 MD simulations on a 2D soft-core mixture, we have examined 
bond breakage events to find large-scale 
 heterogeneities in the changing rates of the amorphous  
 structure, which are enhanced at low  temperatures and  
 surprisingly similar to  the critical fluctuations in Ising systems.   
 We believe that this is a first step in the  investigation of  more complex
situations and 3D cases. 
 We should also clarify  the relationship of our patterns  
 in the $\alpha$ relaxation 
and those by Muranaka and Hiwatari \cite{Muranaka} in the 
$\beta$ relaxation.

We   thank Dr. T. Muranaka,  Professor Y. Hiwatari and 
 Professor K. Kawasaki for helpful discussions.
This work is supported by Grants in Aid for Scientific 
 Research from the Ministry of Education, Science, Sports and Culture.
Calculations have been carried out at the Supercomputer 
 Laboratory, Institute for Chemical Research, Kyoto University 
 and the Computer Center of the Institute for Molecular Science, 
 Okazaki, Japan.

\section*{Figure Captions}

\noindent
Fig.1  Bonds defined at a given time. Their lengths are 
mostly close to $\sigma_{\alpha\beta}$.

\noindent
Fig.2  Temperature dependence of the bond breakage time $\tau_{b}$ 
($\bullet$), where $T^*=k_BT/\epsilon$.
The dotted line is a viewing guide. The vertical arrows   
are estimates for the $\alpha$ relaxation time from the MD data of
 Muranaka and Hiwatari.\cite{Muranaka_private}  \\

\noindent
Fig.3  Snapshots of the broken bonds at  
  $\Gamma_{eff}=1.0$ and $1.3$. The time interval is $0.05\tau_b$, 
so $5\%$ of the initial bonds are broken here. The arrows indicate 
$\xi$. \\ 

\noindent
Fig.4  Broken bond distributions in  two consecutive  time intervals,  
 $[t_0,t_0+ 0.05\tau_b]$ ($\Box$) and 
 $[t_0+0.05\tau_b,t_0+ 0.1\tau_b]$ ($\bullet$),  
at $\Gamma_{eff}=1.4$. The arrows indicate 
$\xi$.\\

\noindent
Fig.5  $S_b(q)$ of the broken  bond density. 
The insert shows $1/S_b(q)$ vs $q^2$, from which $\xi^{-2}$ 
is determined. \\

\noindent
Fig.6  Temperature dependence of the correlation length $\xi$, 
 where $T^*=k_BT/\epsilon$.
 The widths of the time intervals are  $0.05\tau_b$ ($\Diamond$) 
and $0.1\tau_b$ ($\bullet$). The system length is 111.8.

\end{multicols}
\end{document}